\documentclass[11pt]{article}
\usepackage{latexsym}
\usepackage{chapterbib}
\usepackage{graphics,graphicx}
\usepackage{amssymb,amsmath,amsfonts,amstext}
\usepackage{color,xcolor}
\usepackage{hyperref}
\usepackage{geometry}  
\usepackage{braket}
\usepackage{cancel}
\usepackage[utf8]{inputenc}
\usepackage{slashed}
\usepackage{tikz}
\usepackage{accents}
\usepackage{framed}
\usepackage{bm}
\usepackage{dsfont}

\allowdisplaybreaks 

\def\){\right)}
\def\({\left( }
\def\]{\right] }
\def\[{\left[ }

\def\NO{\nonumber}

\newcommand{\be}{\begin{equation}}
\newcommand{\ee}{\end{equation}}

\def\bea{\begin{eqnarray}}
\def\eea{\end{eqnarray}}

\def\bal#1\eal{\begin{align}#1\end{align}}
\def\bala#1\eala{\begin{align*}#1\end{align*}}
\newcommand\numberthis{\addtocounter{equation}{1}\tag{\theequation}}

\def\bald{\begin{aligned}}
\def\eald{\end{aligned}}

\def\bsub{\begin{subequations}}
\def\esub{\end{subequations}}

\def\beqx{\begin{displaymath}}
\def\eeqx{\end{displaymath}}

\newcommand{\bmat}{\left(\begin{array}}
\newcommand{\emat}{\end{array}\right)}

\def\a{\alpha}
\def\b{\beta}
\def\c{\chi}
\def\d{\delta}
\def\e{\epsilon}

\def\g{\gamma}
\def\h{\eta}
\def\j{\psi}
\def\k{\kappa}
\def\l{\lambda}
\def\m{\mu}
\def\n{\nu}
\def\o{\omega}
    
\def\p{\pi}

\def\r{\rho}
\def\s{\sigma}

\def\x{\xi}

\def\S{\Sigma}

\def\ve{\varepsilon}

\def\ca{{\cal A}}

\def\cc{{\cal C}}

\def\cf{{\cal F}}

\def\ch{{\cal H}}

\def\cm{{\cal M}}
\def\cn{{\cal N}}

\def\bo{{\raise-.3ex\hbox{\large$\Box$}}}               
\def\pa{\partial}                                       
\def\face{{\raise.2ex\hbox{$\displaystyle \bigodot$}\mskip-2.2mu \llap {$\ddot
        \smile$}}}                                   
\def\>{\rangle}                                      
\def\<{\langle}                                      

\def\lbar#1{\ensuremath{\overline{#1}}}              

\def\leftrightarrowfill{$\mathsurround=0pt \mathord\leftarrow \mkern-6mu
        \cleaders\hbox{$\mkern-2mu \mathord- \mkern-2mu$}\hfill
        \mkern-6mu \mathord\rightarrow$}        
\def\dvec#1{\vbox{\ialign{##\crcr
        \leftrightarrowfill\crcr\noalign{\kern-1pt\nointerlineskip}
        $\hfil\displaystyle{#1}\hfil$\crcr}}}           

\def\-{\hphantom{-}}
\newcommand{\dd}{\mbox{d}}

\begin{document}
	
	\begin{titlepage}
		
		\pagestyle{empty}

		\vskip1.5in
		
		\begin{center}
			\textbf{\LARGE Super-Weyl Anomaly from Holography and \\ \vskip.5em 
			Rigid Supersymmetry Algebra on Two-Sphere}
		\end{center}
		\vskip0.2in

		\begin{center}
			{\large Ok Song An, Yong Hae Ko, and Sok-Hyon Won}
		\end{center}
		\vskip0.2in
		
		\begin{center}
			{\small Department of Physics, \textbf{Kim Il Sung} University,\\ Ryongnam Dong, TaeSong District, Pyongyang, DPR Korea}\\ \vskip0.1in
		\end{center}
		\vskip0.6in

		\begin{abstract}

Operators transform anomalously under the symmetry in the presence of quantum anomalies. We study this aspect of the super-Weyl anomaly in $\mathcal N=(1,1)$ superconformal field theories (SCFTs), in the context of AdS/CFT. In particular, we carry out holographic renormalization for $(1,1)$ pure AdS$_3$ supergravity that is supposed to be a gravity dual of the $\mathcal N=(1,1)$ SCFT, and derive holographic superconformal Ward identities with corresponding anomalies. We show that the obtained super-Weyl anomaly of the $\mathcal N=(1,1)$ SCFT induces a quantum correction term in the transformation law of the supercurrent under the rigid supersymmetry. We demonstrate that the correction term does not affect the $\mathcal N=(1,1)$ rigid supersymmetry algebra on two-sphere.
		\end{abstract}
		
	\end{titlepage}

	\tableofcontents
	\addtocontents{toc}{\protect\setcounter{tocdepth}{3}}
	\renewcommand{\theequation}{\arabic{section}.\arabic{equation}}




\section{\label{sec:intro}Introduction}
\setcounter{equation}{0}

In recent works \cite{An:2017ihs,Papadimitriou:2017kzw}, 4 dimensional $\cn=1$ superconformal anomalies including the super-Weyl and supersymmetry anomalies were obtained by carrying out supersymmetric holographic renormalization for the AdS$_5$ supergravity and using the AdS/CFT dictionary \cite{Maldacena:1997re,Witten:1998qj,Gubser:1998bc}. In addition, it was shown that these fermionic anomalies make the supercurrent operator transform anomalously under the rigid supersymmetry on curved manifolds and thus add a quantum correction to the rigid supersymmetry algebra. This result gives a partial answer to the issue raised in \cite{Genolini:2016sxe,Genolini:2016ecx} that the supersymmetric Casimir energy \cite{Cassani:2014zwa,Assel:2015nca,Assel:2014paa, Closset:2013vra} of $\cn=1$ supersymmetric gauge theories defined on supersymmetric backgrounds computed via localization technique \cite{Pestun:2007rz,Nekrasov:2002qd} does not agree with the holographic computation, because the localization technique is based on the rigid supersymmetry algebra with no quantum correction term.\footnote{The rigid supersymmetry algebra used in supersymmetric localization is derived from the classical supergravity algebra and thus does not take into account the quantum effect. The result of \cite{An:2017ihs,Papadimitriou:2017kzw} is not a complete answer to the mismatch, since one has yet to verify by means of purely field-theoretic computations if the rigid supersymmetry algebra really receives a quantum correction.}

In this note we would like to carry out an analysis similar to the one in \cite{An:2017ihs,Papadimitriou:2017kzw}, for the AdS$_3$/CFT$_2$ correspondence. That is, we perform supersymmetric holographic renormalization for $(1,1)$ pure AdS$_3$ supergravity to obtain the super-Weyl anomaly of 2 dimensional $\cn=(1,1)$ SCFTs and discuss its physical implications.\footnote{The holographic super-Weyl anomaly for the AdS$_3$/CFT$_2$ correspondence was obtained in \cite{Nishimura:1999gg}, where its physical implications were not discussed.}

Following \cite{An:2017ihs,Papadimitriou:2017kzw,Martelli:2002sp,Papadimitriou:2010as,Papadimitriou:2011qb,Papadimitriou:2016yit}, we first formulate the bulk supergravity in terms of the radial Hamiltonian formalism to obtain the radial Hamiltonian of the bulk theory.\footnote{In \cite{Papadimitriou:2016xxx}, AdS$_3$/CFT$_2$ without supersymmetry was studied by the same approach to reproduce the well-known result that asymptotic charges of pure AdS$_3$ satisfy the Virasoro algebra of 2 dimensional CFT.} The equations of motion for this radial Hamiltonian lead to the first-class constraints that reflect the bulk gauge symmetries. These first-class constraints in turn become the Hamilton-Jacobi (HJ) equations for the Hamilton's principal function $S$, whose solution is just equivalent to the on-shell supergravity action.  One can identify the boundary counterterms with the opposite sign of the divergent part of the asymptotic solution of the HJ equations.

The counterterms allow to define the renormalized canonical momenta, which, according to the AdS/CFT dictionary, correspond to the renormalized one-point functions of the boundary field theory. As in \cite{An:2017ihs,Papadimitriou:2017kzw}, we obtain the Ward identities that reflect Weyl symmetry, diffeomorphism invariance, supersymmetry and super-Weyl symmetry of $\cn=(1,1)$ SCFTs with the supersymmetrized Weyl anomaly and super-Weyl anomaly.

The super-Weyl anomaly itself vanishes on bosonic backgrounds where quantum field theories are usually studied, and thus it seems that the super-Weyl anomaly has no physical meaning. However, this is not the case. Indeed, it turns out that the holographic Ward identities allow us to find the physical implications of the super-Weyl anomaly. First, it follows from the super-Weyl Ward identity that the two-point correlation function of the supercurrent and its supertrace is a contact term. This can be verified in 2 dimensional SCFTs by manipulating the two-point function of the supercurrent operators. Second, the super-Weyl anomaly plays a role of inducing a quantum correction term in the action of the rigid supersymmetry on the supercurrent that can be obtained from the supersymmetry and super-Weyl Ward identities. It is this quantum correction that generates the central term in the anti-commutator of the modes of the supercurrents in the super-Virasoro algebra.

In \cite{An:2017ihs,Papadimitriou:2017kzw} it was shown that the quantum correction in the transformation law of the supercurrents under the rigid supersymmetry leads to a correction term in 4 dimensional $\cn=1$ rigid supersymmetry algebra. On the contrary, in this note we demonstrate that the $\cn=(1,1)$ rigid supersymmetry algebra on two-sphere receives no quantum correction. We emphasize that this phenomenon is special to 2 dimensional manifolds and there is no analogue in 4 dimensions.

The rest of this note is organized as follows. In section \ref{sec:radial-hamiltonian} we formulate the $\cn=(1,1)$ pure AdS$_3$ supergravity action in the radial Hamiltonian formalism, derive the first-class constraints and determine the boundary counterterms to renormalize the on-shell supergravity action and the canonical momenta. In section \ref{sec:holo-ward} we derive the holographic Ward identities and discuss the physical implication of the super-Weyl anomaly. In section \ref{sec:transformations} we obtain the action of the rigid symmetries on the operators in 2 dimensional SCFTs. In section \ref{sec:rigid-susy} we discuss the anomalous transformation of the supercurrent under the rigid supersymmetry and demonstrate that the anomalous term does not affect the $\cn=(1,1)$ rigid supersymmetry algebra on two-sphere. We list some concluding remarks in section \ref{sec:conclusions}. Appendices contain our conventions and some technical details.

\section{Radial Hamiltonian and holographic renormalization}
\setcounter{equation}{0}
\label{sec:radial-hamiltonian}

In this section we apply the supersymmetric holographic renormalization developed in \cite{An:2017ihs,Papadimitriou:2017kzw} to $\cn=(1,1)$ pure AdS$_3$ supergravity.

The bulk action for pure $\cn=(1,1)$ AdS$_3$ supergravity reads \cite{Achucarro:1987vz}
\begin{align*}
S_{\rm bulk}&=\frac{1}{2\k^2}\int_\cm \dd^3x\;\hat{\bm e}\Big(R-\bar\j_{\hat\m}\hat\g^{\hat\m\hat\n\hat\r}\hat\nabla_{\hat\n}\j_{\hat\r}-\bar\j'_{\hat\m}\hat\g^{\hat\m\hat\n\hat\r}\hat\nabla_{\hat\n}\j'_{\hat\r}\\
&\hskip3em-\frac{1}{2L}\bar\j_{\hat\m}\hat\g^{\hat\m\hat\n}\j_{\hat\n}+\frac{1}{2L}\bar\j'_{\hat\m}\hat\g^{\hat\m\hat\n}\j'_{\hat\n}+\frac{2}{L^2}\Big),\numberthis\label{eqn:bulkaction}
\end{align*}
where 4-fermion terms are omitted. Here $\cm$ stands for the three dimensional spacetime in which the supergravity theory is defined. In addition, $\j_{\hat\m}$ and $\j'_{\hat\m}$ refer to the Majorana gravitino of the $(1,0)$ and $(0,1)$ sector, and $\hat\nabla$ denotes the bulk covariant derivative. Finally, $-2/L^2$ is the cosmological constant and $\hat{\bm e}=\det \hat e^{\hat a}_{\hat\m}$.

The supergravity action \eqref{eqn:bulkaction} is invariant under the local transformations
\begin{subequations}\label{eqn:bulk-symmetries}
	\begin{align}
	& \d_{\x,\l,\e,\e'} \hat E^{\hat a}_{\hat\m}=\x^{\hat\n}\pa_{\hat\n}\hat E^{\hat a}_{\hat\m}+\pa_{\hat\m}\x^{\hat\n}\hat E^{\hat a}_{\hat\n}+\l^{\hat a}{}_{\hat b}\hat E^{\hat b}_{\hat\m}+\frac12\bar\e\g^{\hat a}\j_{\hat\m}+\frac12\bar\e'\g^{\hat a}\j'_{\hat\m},\\ & \d_{\x,\l,\e}\j_{\hat\m}=\x^{\hat\n}\pa_{\hat\n}\j_{\hat\m}+(\pa_{\hat\m}\x^{\hat\n})\j_{\hat\n}+\frac14\l^{\hat a\hat b}\g_{\hat a\hat b}\j_{\hat\m}+\hat\nabla_{\hat\m}\e-\frac{1}{2L}\hat\g_{\hat\m}\e,\\
	&\d_{\x,\l,\e}\j'_{\hat\m}=\x^{\hat\n}\pa_{\hat\n}\j'_{\hat\m}+(\pa_{\hat\m}\x^{\hat\n})\j'_{\hat\n}+\frac14\l^{\hat a\hat b}\g_{\hat a\hat b}\j'_{\hat\m}+\hat\nabla_{\hat\m}\e'+\frac{1}{2L}\hat\g_{\hat\m}\e',
	\end{align}
\end{subequations}
up to the boundary terms. Here $\x^{\hat\m}$, $\l^{\hat a\hat b}$, $\e$ and $\e'$ are the parameters of the local diffeomorphism, local Lorentz and local supersymmetry transformations, respectively.

In asymptotically locally AdS spaces (AlAdS) the asymptotic behavior of a spinor field is determined by its mass term and radiality \cite{Kalkkinen:2000uk,Henneaux:1998ch,Arutyunov:1998ve}.
In the case of $\cn=(1,1)$ AdS$_3$ supergravity, $\j_{\m+}$ and $\j'_{\m-}$ are more dominant than $\j_{\m_-}$ and $\j'_{\m+}$ near the boundary and therefore it is natural to take $\j_{\m+}$ and $\j'_{\m-}$ as the source fields of the boundary field theory. This implies that the Dirichlet boundary condition is imposed for $\j_{\m+}$ and $\j'_{\m-}$ in the bulk supergravity.

It then follows that in order to have a well-posed variational problem we should add the super-Gibbons-Hawking boundary terms \cite{Gibbons:1976ue,Nishimura:1999gg}
\begin{align}
	S_{\rm sGH}&=\frac{1}{2\k^2}\int_{\pa\cm}\dd^2x\sqrt{-g}\;\Big(2K-\frac12\ve^{\m\n}\bar\j_\m\g_*\j_\n+\frac12\ve^{\m\n}\bar\j'_\m\g_*\j'_\n\Big),
\end{align}
to the supergravity action \eqref{eqn:bulkaction}, where $K\equiv g^{\m\n}K_{\m\n}$ and 
\begin{equation}
K_{\m\n}\equiv \frac{1}{2N}(\dot g_{\m\n}-\nabla_\m N_\n-\nabla_\n N_\m)
\end{equation}
is the extrinsic curvature.

In order to formulate the supergravity theory in terms of the radial Hamiltonian formalism, we begin with radial-decomposition of the full action
\begin{equation}
	S_{\rm full}=S_{\rm bulk}+S_{\rm sGH}=\int\dd r\;L,
\end{equation}
where the bulk is regarded as a foliation of the radial slices $\S_r$ and $L$ is the so-called radial Lagrangian, which is given by
\begin{align*}
	L&=\frac{1}{2\k^2}\int_{\S_r}\dd^2x\;\sqrt{-g}\Big\{N\big(R[g]+K^2-K^{\m\n}K_{\m\n}+\frac{2}{L^2}\big)+2\ve^{\m\n}\dot{\lbar{\j_{\m+}}}\j_{\n-}+2\ve^{\m\n}\dot{\lbar{\j'_{\m-}}}\j'_{\n+}\\
	&\hskip4em+\ve^{\m\n}(\bar\j_r-N^\l\bar\j_\l)K_{\r\m}\g^\r\g_*\j_\n+\frac14\ve^{\m\n}\dot E_{a\r}E^\r_b\bar\j_\m\g^{ab}\j_\n-\frac12\ve^{\m\n}\pa_\r N\bar\j_\m\g_*\g^\r\j_\n\\
	& \hskip4em+\frac14\ve^{\m\n}\nabla_\r N_\l\bar\j_\m\g^{\r\l}\j_\n-\frac1L(\bar\j_r-N^\l\bar\j_\l)\g_*\g^\m\j_\m -\frac{N}{2L}\bar\j_\m\g^{\m\n}\j_\n+2\ve^{\m\n}\bar\j_r\nabla_\m\j_\n\\
	&\hskip4em+\ve^{\m\n}(\bar\j'_r-N^\l\bar\j'_\l)K_{\r\m}\g^\r\g_*\j'_\n+\frac14\ve^{\m\n}\dot E_{a\r}E^\r_b\bar\j'_\m\g^{ab}\j'_\n-\frac12\ve^{\m\n}\pa_\r N\bar\j'_\m\g_*\g^\r\j'_\n\\
	& \hskip4em+2\ve^{\m\n}\bar\j'_r\nabla_\m\j'_\n+\frac14\ve^{\m\n}\nabla_\r N_\l\bar\j'_\m\g^{\r\l}\j'_\n+\frac1L(\bar\j'_r-N^\l\bar\j'_\l)\g_*\g^\m\j'_\m +\frac{N}{2L}\bar\j'_\m\g^{\m\n}\j'_\n \Big\}.\numberthis
\end{align*}
The radial Lagrangian allows to define the canonical conjugate momenta. The conjugate momenta for $\j_r$, $\j'_r$, $N^\m$ and $N$ vanish identically, since  their radial derivatives  are not involved in the radial Lagrangian $L$. In fact, as is well-known, they are the Lagrange multipliers that parameterize the gauge transformations of the supergravity theory.

The conjugate momenta for $E^a_\m$, $\j_{\m+}$ and $\j'_{\m-}$ are given as
\begin{subequations}
		\begin{align*}
	\p^\m_a&\equiv \frac{\d L}{\d \dot E^a_\m}=\frac{\sqrt{-g}}{2\k^2}(\d^\m_\n E_{\r a}+\d^\m_\r E_{\n a})\Big[g^{\n\r}K-K^{\n\r} +\frac{1}{2N}[\ve^{\n\s}(\bar\j_r-N^\l\bar\j_\l)\g^\r\g_*\j_\s]+\\
	&\hskip3em+\frac{1}{2N}[\ve^{\n\s}(\bar\j'_r-N^\l\bar\j'_\l)\g^\r\g_*\j'_\s]\Big]+\frac{\sqrt{-g}}{8\k^2}E^{b\m}\ve^{\r\s}\bar\j_\r\g_{ab}\j_\s+\frac{\sqrt{-g}}{8\k^2}E^{b\m}\ve^{\r\s}\bar\j'_\r\g_{ab}\j'_\s,\numberthis\\
	\p_\j^\m&\equiv\frac{\d L}{\d \dot{\lbar{\j_{\m+}}}}=\frac{\sqrt{-g}}{\k^2}\ve^{\m\n}\j_{\n-},\numberthis\\
	\p_{\j'}^\m&\equiv\frac{\d L}{\d \dot{\lbar{\j'_{\m-}}}}=\frac{\sqrt{-g}}{\k^2}\ve^{\m\n}\j'_{\n+},\numberthis
	\end{align*}
\end{subequations}
From the definition of the conjugate momenta $\p_a^\m$, we obtain a primary first-class constraint
\begin{equation}\label{eqn:local-lorentz}
	\frac{\k^2}{\sqrt{-g}}(E^{a\m}\p_a^\n-E^{a\n}\p_a^\m)=\frac14\ve^{\r\s}(\bar\j_\r\g^{\m\n}\j_\s+\bar\j'_\r\g^{\m\n}\j'_\s),
\end{equation}
which reflects the local Lorentz invariance of the bulk supergravity theory.

Now that the conjugate momenta are given, we can obtain the radial Hamiltonian by a Legendre transformation
\begin{align}
	H&=\int_{\S_r}\dd^2x\;(\p^\m_a\dot E_\m^a+\dot{\lbar{\j_{\m+}}}\p_\j^\m+\dot{\lbar{\j'_{\m-}}}\p_{\j'}^\m-L)\NO\\
	&=\int_{\S_r}\dd^2x\;[N\ch+N^\m\ch_\m+(\bar\j_r-N^\m\bar\j_\m)\cf +(\bar\j'_r-N^\m\bar\j'_\m)\cf'],\label{eqn:radial-hamiltonian}
\end{align} 
where
\begin{subequations}
	\begin{align*}
	\ch&=\frac{\k^2}{2\sqrt{-g}}(E^a_\m E^b_\n-E^a_\n E^b_\m)\p_a^\m\p_b^\n+\frac{1}{2L} \bar\j_{\m+}\p_\j^\m+\frac{1}{2L} \bar\j'_{\m-}\p_{\j'}^\m+\\
	&\hskip1em-\frac{\k^2}{4\sqrt{-g}}\ve_{\m\n}\nabla_\r(\bar\p_\j^\m\g^\r\p_\j^\n)+\frac{\k^2}{4\sqrt{-g}}\ve_{\m\n}\nabla_\r(\bar\p_{\j'}^\m\g^\r\p_{\j'}^\n)-\\
	&\hskip1em-\frac{\sqrt{-g}}{2\k^2}\Big(R[g]+\frac{2}{L^2}-\frac12\ve^{\m\n}\nabla_\r(\bar\j_{\m+}\g^\r\j_{\n+})+\frac12\ve^{\m\n}\nabla_\r(\bar\j'_{\m-}\g^\r\j'_{\n-})\Big),\numberthis\\
	\ch_\m&=-E^a_\m\nabla_\n\p_a^\n+(\nabla_\m\bar\j_{\n-})\p_\j^\n-\nabla_\n(\bar\j_{\m-}\p_\j^\n)+(\nabla_\m\bar\j'_{\n-})\p_{\j'}^\n-\nabla_\n(\bar\j'_{\m-}\p_{\j'}^\n),\numberthis \\
	\cf_+&=\frac{\k^2}{4\sqrt{-g}}(2g_{\m\n}\g_\r-g_{\m\r}\g_\n-g_{\n\r}\g_\m)\p_\j^\r E^{a\n}\p_a^\m+\frac{1}{2L}\g_\m\p_\j^\m-\frac{\sqrt{-g}}{\k^2}\ve^{\m\n}\nabla_\m\j_{\n+},\numberthis\\
	\cf_-&=\frac12\g_{(\m}\j_{\n+)}E^{a\n}\p_a^\m-\frac{\sqrt{-g}}{2\k^2 L}\g^\m\j_{\m+}-\nabla_\m\p_\j^\m\numberthis ,\\
	\cf'_+&=\frac12\g_{(\m}\j'_{\n-)}E^{a\n}\p_a^\m-\frac{\sqrt{-g}}{2\k^2 L}\g^\m\j'_{\m-}-\nabla_\m\p_{\j'}^\m,\numberthis\\
	\cf'_-&=-\frac{\k^2}{4\sqrt{-g}}(g_{\m\n}g_{\r\l}-g_{\m\r}g_{\n\l})(\g^\l\p_{\j'}^\r+\g^\r\p_{\j'}^\l)E^{a\n}\p_a^\m-\frac{1}{2L}\g_\m\p_{\j'}^\m-\frac{\sqrt{-g}}{\k^2}\ve^{\m\n}\nabla_\m\j'_{\n-},\numberthis
	\end{align*}
\end{subequations}
and $\cf_\pm \equiv (1\pm\g_*)\cf/2$ and $\cf'_\pm \equiv (1\pm\g_*)\cf'/2$.

Since the conjugate momenta for $N$, $N^\m$ and $\j_r$ vanish by construction, their Hamiltonian equations of motion become
\begin{equation}\label{eqn:first-class-constraint}
	\ch=\ch_\m=\cf_-=\cf_+=\cf'_-=\cf'_+=0,
\end{equation}
which are in fact the first class constraints that generate the scaling, diffeomorphism, super-Weyl and supersymmetry transformations respectively on the radial slice $\S_r$. In particular, $N$, $N^\m$, $\j_r$ and $\j_r'$ are gauge degrees of freedom in the bulk supergravity and we are free to choose a certain gauge for them. In this work we choose the so-called Fefferman-Graham gauge, namely
\begin{equation}\label{eqn:FG-gauge}
	N=1,\quad N^\m=\j_r=\j_r'=0.
\end{equation}

The Hamilton-Jacobi (HJ) equations for the Hamilton's principal function $S$ are obtained by substituting
\begin{equation}
\p_a^\m=\frac{\d S}{\d E^a_\m},\quad \p_\j^\m=\frac{\d S}{\d \bar\j_{\m+}},\quad \p_{\j'}^\m=\frac{\d S}{\d\bar\j'_{\m-}},
\end{equation}
into the first class constraints \eqref{eqn:first-class-constraint}. Now we can determine the boundary counterterms by solving the HJ equations recursively with respect to the dilatation operator
\begin{equation}
\d_D\equiv \int_{\S_r}\dd^2x \(e^a_\m \frac{\d}{\d e^a_\m}+\frac12\bar\c_\m\frac{\d}{\d\bar\c_\m}+\frac12\bar\c'_\m\frac{\d}{\d\bar\c'_\m} \),
\end{equation}
where $e^a_\m$, $\c_\m$ and $\c'_\m$ are induced zweibein and $(1,0)$ and $(0,1)$ gravitino fields on the radial slice $\S_r$, respectively. The solution of HJ equations can be written as
\begin{equation}
S=S_{(0)}+\tilde S_{(2)}+S_{(2)}+\cdots ,\quad \d_D S_{(0)}=2S_{(0)},\;\d_D \tilde S_{(2)}=0,
\end{equation}
where $S_{(0)}$ is quadratically divergent, and $\tilde S_{(2)}$ is the logarithmically divergent term that essentially indicates the Weyl anomaly of the dual boundary field theory. $S_{(2)}$ is finite and actually equivalent to the renormalized on-shell action.

In the case of pure AdS$_3$ supergravity, the divergent terms of the asymptotic HJ solution are simply given by
\begin{align}
S_{\rm div}&=\frac{1}{\k^2}\int_{\S_r}\dd^2x\sqrt{-g}\Big[\frac 1L-\frac L2 \log e^{-\frac{2r}{L}}\Big(R[g]-\frac12\ve^{\m\n}\nabla_\r(\bar\j_{\m+}\g^\r\j_{\n+}-\bar\j'_{\m-}\g^\r\j'_{\n-})\Big)\Big].\label{eqn:div-action}
\end{align}
The counterterms are simply identified with $-S_{\rm div}$, namely
\begin{equation}
S_{\rm ct}=-S_{\rm div},
\end{equation}
which removes the long-distance divergence of the bulk supergravity on-shell action. Therefore, the renormalized on-shell action is defined as
\begin{equation}
	S_{\rm ren}=\lim_{r\to\infty}(S_{reg}+S_{\rm ct}),
\end{equation}
where $S_{reg}$ is the regularized on-shell action. Note that the logarithmic term in $S_{\rm ct}$ explicitly depends on the radial coordinate $r$ and thus implies that the diffeomorphism invariance of the bulk supergravity theory along the radial direction is broken. In the AdS/CFT context this corresponds to the breakdown of Weyl invariance of the dual CFT on the boundary. In particular, the logarithmic term in $S_{\rm ct}$ becomes the Weyl anomaly of the dual CFT.

The asymptotic behavior of the source fields can be determined from the asymptotic HJ solution $S_{\rm div}$. In fact, the Hamiltonian equations of motion lead us to the flow equations
\begin{subequations}
	\begin{align}
	\dot E^a_\m&=\frac{\d H}{\d \p_a^\m}=\frac{\k^2}{\sqrt{-g}}(E^a_\m E^b_\n-E^a_\n E^b_\m)\p_b^\n,\\
	\dot \j_{\m+}&=\frac{\d H}{\d \bar\p_\j^\m}=\frac{1}{2L}\j_{\m+},\\
	\dot \j'_{\m-}&=\frac{\d H}{\d \bar\p_{\j'}^\m}=\frac{1}{2L}\j'_{\m-}.
	\end{align}
\end{subequations}
From these flow equations, we see that the asymptotic behavior of the source fields are
\begin{subequations}
	\begin{align}
	& E^a_\m(r,x)\sim e^{\frac rL}e^a_{\m}(x),\\
	& \j_{\m+}(r,x)\sim e^{\frac{r}{2L}} \c_\m(x),\\
	&\j'_{\m-}(r,x)\sim e^{\frac{r}{2L}}\c'_{\m}(x),
	\end{align} 
\end{subequations}
where $e^a_\m(x)$, $\c_\m(x)$ and $\c'_\m(x)$ can be regarded as the source fields of the boundary CFT that couple to the stress-energy tensor and supercurrent operator respectively.

The renormalized canonical momenta are then defined as
\begin{subequations}
	\begin{align}
	&	\hat \p_a^\m\equiv \lim_{r \to +\infty}e^{\frac rL}\(\p_a^\m+\frac{\d S_{\rm ct}}{\d E^a_\m}\),\\
	& \hat\p_\j^\m\equiv \lim_{r\to+\infty}e^{\frac{r}{2L}}\(\p_\j^\m+\frac{\d S_{\rm ct}}{\d \bar\j_{\m+}}\),\\
	& \hat\p_{\j'}^\m\equiv \lim_{r\to+\infty}e^{\frac{r}{2L}}\(\p_{\j'}^\m+\frac{\d S_{\rm ct}}{\d \bar\j'_{\m-}}\).
	\end{align}
\end{subequations}
The variation of the renormalized on-shell action becomes
\begin{equation}
	\d S_{\rm ren}=\int_{\pa\cm}\dd^2x\(\hat\p_a^\m\d e^a_\m+\d\bar\c_\m\hat\p_\j^\m+\d\bar\c_\m'\hat\p_{\j'}^\m  \),
\end{equation}
which implies that the variational problem is now well-posed under the Dirichlet boundary condition for the source fields.

\section{Holographic Ward identities and anomalies}
\label{sec:holo-ward}
\setcounter{equation}{0}

The above renormalized canonical momenta correspond to renormalized one-point functions of the stress-energy tensor and the supercurrent operator of the dual field theory by the AdS/CFT dictionary, namely
\begin{subequations}
\begin{align}
& \braket{T_a^\m} \equiv -2\p\bm e^{-1}\hat\p_a^\m,\quad T^{\m\n}\equiv e^{a\m}T_a^\n,\\
& \braket{T_F^\m}\equiv\begin{pmatrix}
0 \\ S^\m
\end{pmatrix}\equiv 2\p\bm e^{-1} \hat\p_\j^\m,\\
& \braket{T'^\m_F}\equiv\begin{pmatrix}
 S'^\m \\ 0
\end{pmatrix}\equiv 2\p\bm e^{-1} \hat\p_{\j'}^\m,
\end{align}
\end{subequations}
where $\bm e=\det e^a_\m$.

Using this mapping, the first class constraints \eqref{eqn:first-class-constraint} and \eqref{eqn:local-lorentz} are converted into the quantum Ward identities that relate the above one-point functions, namely
\begin{subequations}\label{eqn:ward-identities}
\begin{align}
& e^a_\m \braket{T_a^\m}-\frac{1}{2}\bar \c_\m\braket{T_F^\m}-\frac12\bar\c'_\m\braket{T'^\m_F}=\ca_{\rm w},\label{eqn:ward-trace}\\
& e^a_\m\nabla_\n \braket{T^\n_a}+(\nabla_\m\bar\c_\n)\braket{T_F^\n}-\nabla_\n(\bar\c_\m \braket{T_F^\n})+(\nabla_\m\bar\c'_\n)\braket{T_F'^\n}-\nabla_\n(\bar\c'_\m \braket{T_F'^\n})=0,\label{eqn:ward-diffeo}\\
& \nabla_\m \braket{T_F^\m}+\frac12\g_{(\m}\c_{\n)}e^{a\n}\braket{T_a^\m}=0,\quad \nabla_\m \braket{T_F'^\m}+\frac12\g_{(\m}\c'_{\n)}e^{a\n}\braket{T_a^\m}=0,\label{eqn:ward-supersymmetry}\\
& \g^\m \braket{T_{F\m}}=\ca_{\rm sw},\quad \g^\m \braket{T'_{F\m}}=\ca'_{\rm sw},\label{eqn:ward-supertrace}\\
& e_{a\m}\braket{T^\m_b}-e_{b\m}\braket{T^\m_a}-\frac12\bar\c_\m\g_{ab}\braket{T_F^\m}-\frac12\bar\c'_\m\g_{ab}\braket{T'^\m_F}=0,\label{eqn:ward-lorentz}
\end{align}
\end{subequations}
where the Weyl and super-Weyl anomalies are respectively given as
\begin{subequations}
	\begin{align}
	\ca_{\rm w} &=-2\p\frac{L}{2\k^2}\Big(R[e]-\frac12\ve^{\m\n}\nabla_\r(\bar\c_\m\g^\r\c_\n)+\frac12\ve^{\m\n}\nabla_\r(\bar\c'_\m\g^\r\c'_\n)\Big),\label{eqn:anomaly-trace}\\
	\ca_{\rm sw}&=2\p\frac{L}{\k^2}\ve^{\m\n}\nabla_\m\c_\n,\quad \ca'_{\rm sw}=-2\p\frac{L}{\k^2}\ve^{\m\n}\nabla_\m\c'_\n.\label{eqn:anomaly-superweyl}
	\end{align}
\end{subequations}
The Ward identities \eqref{eqn:ward-identities} reflect the conformal symmetry, diffeomorphism invariance, supersymmetry, super-Weyl symmetry and local Lorentz invariance of the dual SCFT. Notice that the local Lorentz invariance Ward identity \eqref{eqn:ward-lorentz} implies that $T^{\m\n}$ is no longer a symmetric tensor due to the gravitino sources.

The holographic super-Weyl anomalies \eqref{eqn:anomaly-superweyl} satisfy Wess-Zumino consistency condition, and precisely coincide with the result of \cite{Nishimura:1999gg}, where the anomalies were obtained by using the Fefferman-Graham expansion of the bulk fields and explicitly evaluating the variation of the total supergravity action together with the counterterms. The bosonic part of the holographic Weyl anomaly \eqref{eqn:anomaly-trace} is well-known since the work by Henningson and Skenderis \cite{Henningson:1998gx}, whereas the fermionic part of \eqref{eqn:anomaly-trace} is novel and is the supersymmetry partner of the bosonic part. In fact, the Weyl anomaly \eqref{eqn:anomaly-trace} is superconformal invariant, which can be easily verified.

By comparing with the Weyl anomaly in 2D CFT
\begin{equation}\label{eqn:anomaly-conformal}
	\braket{T^\m_\m}=-\frac{c}{12}R
\end{equation}
one can find that the bulk parameters $L$ and $\k^2$ are related to the central charge of the boundary CFT by \cite{Brown:1986nw}
\begin{equation}\label{eqn:bulk-boundary-constant}
	2\p\frac{L}{\k^2}=\frac c6.
\end{equation}

The super-Weyl anomalies \eqref{eqn:anomaly-superweyl} deserve the further comment. As in AdS$_5$/CFT$_4$ \cite{An:2017ihs,Papadimitriou:2017kzw}, the super-Weyl anomalies $\ca_{\rm sw}$ and $\ca_{\rm sw}'$ depend on the gravitino source field $\c_\m$. Since fermionic backgrounds are usually turned off in the quantum field theory, at first sight it seems that the super-Weyl anomalies have no physical meaning. However, by differentiating \eqref{eqn:ward-supertrace} with respect to the gravitino source field, we find that they contribute the contact terms to the two-point function of the supercurrent operators, namely
\begin{align}\label{eqn:supercurrent-two-point}
	& \braket{\bar T_{F\m}(x)\g^\m T_F^\n(y)}=2\p \frac c6\ve^{\m\n}\nabla_\m\d^2(x,y),\quad \braket{\bar T'_{F\m}(x)\g^\m T_F'^\n(y)}=-2\p \frac c6\ve^{\m\n}\nabla_\m\d^2(x,y),
\end{align}
where $\d^2(x,y)\equiv \d^2(x-y)/\sqrt{-h}$ is the covariant 2 dimensional Dirac delta function. In fact, this is a common feature of the anomalies. For instance, the conformal anomaly \eqref{eqn:anomaly-conformal} vanishes on the flat metric, while it contributes the contact term to the two point function of $T^\m_\m$, namely (see e.g. (4.97) in \cite{Blumenhagen:2013fgp})
\begin{equation}
	\braket{T_{z\bar z}(z)T_{w\bar w}(w)}=-c\frac{\p}{6}\pa_z\bar\pa_{\bar z}\d^{(2)}(z-w),
\end{equation}
where $z$ and $\bar z$ are the coordinates of the complex plane with the metric $\dd s^2=\dd z\dd \bar z$.

One can confirm that the two-point function \eqref{eqn:supercurrent-two-point} exactly matches the 2D SCFT result. First, on the light-cone flat metric \eqref{eqn:light-cone-metric}, the two-point function \eqref{eqn:supercurrent-two-point} becomes
\begin{equation}\label{eqn:Szbar-Sw}
	\braket{S_{+}(x^-_1)S_-(x^-_2)}=-2\p\frac{c}{12}\pa_{x_1^-}\d^{(2)}(x_1^--x_2^-).
\end{equation}
On the 2D SCFT side, we begin with the two-point function of the supercurrent
\begin{equation}
\braket{S_{z}(z)S_{w}(w)}=\frac{\frac c6}{(z-w)^3}.
\end{equation}
Taking a functional derivative with respect to $\bar z$ and using the conservation of the supercurrent $\pa_{\bar z}S_z+\pa_z S_{\bar z}=0$, we find
\begin{equation}
\pa_z\braket{S_{\bar z}(z)S_{w}(w)}=-\frac c6\pa_{\bar z}\frac{1}{(z-w)^3}=-\frac{c}{12}\pa_{\bar z}\pa_z^2\frac{1}{z-w},
\end{equation}
which is consistent with \eqref{eqn:Szbar-Sw}, taking into account that $\pa_{\bar z}(1/(z-w))=2\p\d^{(2)}(z-w)$.\footnote{Since the metric signature of the bulk supergravity is Minkowskian, there is an ambiguity of the sign and a factor $i$ in the map from the supercurrent operator $S_\pm$ defined in terms of the holographic dictionary to the supercurrent $S_{z, \bar z}$ of the SCFT on the complex plane. However, the above analysis is enough to show the physical relevance of the super-Weyl anomaly.}

Therefore we conclude that the super-Weyl anomaly \eqref{eqn:anomaly-superweyl} indeed implies the breakdown of the classical super-Weyl invariance of the SCFTs even on flat backgrounds.

\section{Transformation of the operators}
\setcounter{equation}{0}
\label{sec:transformations}

Another aspect of the above anomalies is that they give rise to the anomalous terms in the transformation of the operators. Indeed, the trace anomaly in 2D CFT makes the stress-energy tensor transform anomalously under the conformal transformation. In the holography context this was shown in \cite{deHaro:2000vlm} by using Fefferman-Graham expansion of the bulk metric. As shown in \cite{Cvetic:2016eiv}, one can reach the same conclusion easily by using the first class constraints \eqref{eqn:first-class-constraint} and the Poisson bracket. Following \cite{Cvetic:2016eiv,An:2017ihs,Papadimitriou:2017kzw}, we would like to show that the supercurrent operator also transforms anomalously under the rigid supersymmetry.

To this aim, we first need to understand the symmetries of the boundary field theory. There are still residual bulk gauge transformations \eqref{eqn:bulk-symmetries} that does not change the Fefferman-Graham gauge \eqref{eqn:FG-gauge}. These are known as Penrose-Brown-Henneaux (PBH) transformations, under which the variation of the source fields is given as follows \cite{An:2017ihs,Papadimitriou:2017kzw}.
\begin{subequations}\label{eqn:PBH-source}
	\begin{align}
	&	\d_{\s,\x,\l,\e,\h} e^a_\m=\frac{\s}{L} e^a_\m+\x^\n\pa_\n e^a_\m+(\pa_\m\x^\n)e^a_\n-\l^a{}_be^b_\m+\frac12\bar\e\g^a\c_\m+\frac12\bar\e'\g^a\c'_\m,\\
	& \d_{\s,\x,\l,\e,\h}\c_\m=\frac{\s}{2L}\c_\m+\x^\n\pa_\n\c_\m+\pa_\m\x^\n\c_\n-\frac14\l^{ab}\g_{ab}\c_\m+\nabla_\m\e-\frac 1L\g_\m\h,\\
	& \d_{\s,\x,\l,\e',\h'}\c'_\m=\frac{\s}{2L}\c'_\m+\x^\n\pa_\n\c'_\m+\pa_\m\x^\n\c'_\n-\frac14\l^{ab}\g_{ab}\c'_\m+\nabla_\m\e'+\frac 1L\g_\m\h',
	\end{align}
\end{subequations}
where $\s$, $\x^\m$, $\l^a{}_b$, $\e$, $\e'$, $\h$, and $\h'$ are arbitrary functions of the boundary coordinates and parameterize respectively the Weyl, diffeomorphism, local Lorentz, supersymmetry and super-Weyl transformations of the boundary field theory in the context of the AdS/CFT.

The action of the PBH transformation on the canonical variables can be understood by using the Poisson braket, which is defined as \cite{Henneaux:1992ig}
\begin{align}
	\{ A,B\}&=\int\dd^2x\Big[\frac{\d A}{\d e^a_\m}\frac{\d B}{\d \hat \p^\m_a}-\frac{\d B}{\d e^a_\m}\frac{\d A}{\d \hat \p_a^\m}+(-)^{\ve_A}\big(\frac{\d A}{\d \bar\c_\m}\frac{\d B}{\d\hat\p_\j^\m}+\frac{\d A}{\d\hat\p_\j^\m}\frac{\d B}{\d\bar\c_\m}+\frac{\d A}{\d \bar\c'_\m}\frac{\d B}{\d\hat\p_{\j'}^\m}+\frac{\d A}{\d\hat\p_{\j'}^\m}\frac{\d B}{\d\bar\c'_\m}\big)\Big],
\end{align}
where $\ve_A$ is the Grassmann parity of $A$. One can readily see that  the constraint function $C[\s,\x,\e,\h,\e',\h',\l]$ defined as
\begin{align*}
C[\s,\x,\e,\h,\e',\h',\l]&=\int_{\pa\cm}\dd^2 x\Big\{ -\frac{\s}{L}\Big(e^a_\m\hat\p_a^\m+\frac12\bar\c_\m\hat \p_\j^\m+\frac12\bar\c'_\m\hat \p_{\j'}^\m+\frac{\sqrt{-h}}{2\p}\ca_{\rm w}\Big)\\
&\hskip1em+\x^\m[e^a_\m\nabla_\n\hat\p_a^\n-(\nabla_\m\bar\c_\n)\hat\p_\j^\n+\nabla_\n(\bar\c_\m\hat\p_\j^\n)-(\nabla_\m\bar\c'_\n)\hat\p_{\j'}^\n+\nabla_\n(\bar\c'_\m\hat\p_{\j'}^\n)]\\
&\hskip1em+ \bar\e[\nabla_\m\hat\p_\j^\m-\frac14(\g_\m\c_\n+\g_\n\c_\m)e^{a\n}\hat\p_a^\m]-\frac{\bar\h}{L}\Big(\g_\m\hat\p_\j^\m-\frac{\sqrt{-h}}{2\p}\ca_{\rm sw}\Big)\\
&\hskip1em+ \bar\e[\nabla_\m\hat\p_{\j'}^\m-\frac14(\g_\m\c'_\n+\g_\n\c'_\m)e^{a\n}\hat\p_a^\m]+\frac{\bar\h}{L}\Big(\g_\m\hat\p_{\j'}^\m-\frac{\sqrt{-h}}{2\p}\ca'_{\rm sw}\Big)\\
&\hskip1em+\frac12\l^{ab}(-e_{a\m}\hat\p_b^\m+e_{b\m}\hat\p_a^\m-\frac12\bar\c_\m\g_{ab}\hat\p_\j^\m-\frac12\bar\c'_\m\g_{ab}\hat\p_{\j'}^\m)\Big\},\label{eqn:constraint-function}\numberthis
\end{align*}
which vanishes on the solution space of the bulk supergravity, really generates PBH transformations for the canonical variables by the Poisson bracket. For instance, the Poisson bracket of the constraint function \eqref{eqn:constraint-function} and the source fields becomes
\begin{subequations}
	\begin{align*}
	\d_{\s,\x,\e,\h,\e',\h',\l} e^a_\m&=\{C[\s,\x,\e,\h,\e',\h',\l],e^a_\m\}\\
	&=\frac{\s}{L}e^a_\m+e^a_\n\nabla_\m\x^\n-\l^a{}_be^b_\m+\frac14\bar\e(\g_\m\c_\n+\g_\n\c_\m)e^{a\n}++\frac14\bar\e'(\g_\m\c'_\n+\g_\n\c'_\m)e^{a\n},\label{eqn:PBH-e}\numberthis\\
	\d_{\s,\x,\e,\h,\l} \bar\c_\m&=\{C[\s,\x,\e,\h,\l],\bar\c_\m\}\\
	&=\frac{\s}{2L}\bar\c_\m+\x^\n\nabla_\n\bar\c_\m+(\nabla_\m\x^\n)\bar\c_\n+\nabla_\m\bar\e+\frac 1L\bar\h\g_\m+\frac14\l^{ab}\bar\c_\m\g_{ab},\label{eqn:PBH-chi}\numberthis\\
	\d_{\s,\x,\e',\h',\l}\bar\c'_\m&=\{C[\s,\x,\e',\h',\l],\bar\c'_\m\}\\
	&=\frac{\s}{2L}\bar\c'_\m+\x^\n\nabla_\n\bar\c'_\m+(\nabla_\m\x^\n)\bar\c'_\n+\nabla_\m\bar\e'-\frac 1L\bar\h'\g_\m+\frac14\l^{ab}\bar\c'_\m\g_{ab},\label{eqn:PBH-chi-prime}\numberthis
	\end{align*}
\end{subequations}
which is identified with the PBH transformation  \eqref{eqn:PBH-source} of the source fields in terms of the map
\begin{equation}
	\l_{ab}\to\l_{ab}-\x^\m\o_{\m ab}-\frac14\e_{ab}(\bar\e\g^\r\c_\r+\bar\e'\g^\r\c'_\r).
\end{equation}

The transformation of the renormalized canonical momenta is also determined by the Poisson bracket with the constraint function $C[\s,\x,\e,\h,\e',\h',\l]$. First, considering only the bosonic sector, the stress-energy tensor $T_{\m\n}$ becomes symmetric and its variation under the PBH transformation is obtained by
\begin{align}
\d_{\s,\x}T_{\m\n}&=\{C[\s,\x],-2\p e^{-1}e^a_\m\hat\p_a^\r h_{\n\r}\}\NO \\
&=\x^\l\nabla_\l T_{\m\n}+\nabla_\m\x^\l T_{\l\n}+\nabla_\n\x^\l T_{\m\l}-\frac{2\p}{\k^2}\(\frac12\s h_{\m\n}R-\s R_{\m\n}+\nabla_\m\nabla_\n\s-\Box\s h_{\m\n}\),\label{eqn:PBH-stress}
\end{align}
where $h_{\m\n}\equiv e^a_\m e_{a\n}$ is the induced metric on the boundary. For the supercurrent operators, we compute the Poisson bracket and evaluate it on bosonic backgrounds (i.e. the gravitino sources are set to zero), namely 
\begin{subequations}\label{eqn:PBH-supercurrent}
	\begin{align}
 &\d_{\s,\x,\e,\h,\l} T_{F\m}=\{C[\s,\x,\e,\h,\l],2\p e^{-1}\hat\p_\j^\n h_{\m\n}\}\NO \\
 &\hskip2em =-\frac{\s}{2L}T_{F\m}+\x^\n\nabla_\n T_{F\m}+T_{F\n}\nabla_\m\x^\n-\frac14\l^{ab}\g_{ab}T_{F\m}-\frac14\g^\n\e(T_{\m\n}+T_{\n\m})+\frac{2\p}{\k^2}\ve_{\m\n}\nabla^\n\h,\label{eqn:PBH-supercurrent-L}\\
	&\d_{\s,\x,\e',\h',\l} T'_{F\m}=\{C[\s,\x,\e',\h',\l],2\p e^{-1}\hat\p_{\j'}^\n h_{\m\n}\}\NO \\
	&\hskip2em =-\frac{\s}{2L}T'_{F\m}+\x^\n\nabla_\n T'_{F\m}+T'_{F\n}\nabla_\m\x^\n-\frac14\l^{ab}\g_{ab}T'_{F\m}-\frac14\g^\n\e'(T_{\m\n}+T_{\n\m})+\frac{2\p}{\k^2}\ve_{\m\n}\nabla^\n\h'.\label{eqn:PBH-supercurrent-R}
	\end{align}
\end{subequations}

\subsection*{Asymptotic symmetries}

The PBH transformations that leave the source fields intact, i.e.
\begin{equation}\label{eqn:asymptotic-symmetry-condition}
\d_{\s,\x,\e,\h,\e',\h',\l} e^a_\m=0,\quad \d_{\s,\x,\e,\h,\l} \c_\m=0,\quad \d_{\s,\x,\e',\h',\l} \c'_\m=0
\end{equation}
correspond to asymptotic symmetries of the bulk supergravity, or rigid symmetries of the boundary field theory in the AdS/CFT framework.
It follows from \eqref{eqn:PBH-e} and \eqref{eqn:PBH-chi} that on the bosonic background the condition \eqref{eqn:asymptotic-symmetry-condition} becomes
\begin{align*}
&\frac{\s}{L}e^a_\m+e^a_\n\nabla_\m\x^\n-\l^a{}_be^b_\m=0,\\
&\nabla_\m\e-\frac1L\g_\m\h=0,\quad \nabla_\m\e'+\frac1L\g_\m\h'=0,
\end{align*}
or
\begin{subequations}\label{eqn:conformal-killing-condition}
\begin{align}
&\nabla_\m\x_\n+\nabla_\n\x_\m=h_{\m\n}\nabla_\r\x^\r,\quad \s=-\frac L2\nabla_\m\x^\m,\quad \l_{ab}=\frac12\e_{ab}\ve^{\m\n}\nabla_\m\x_\n\label{eqn:ckv}\\
& \h=\frac L2\g^\m\nabla_\m\e,\quad \g^\n\g_\m\nabla_\n\e=0,\quad \h'=-\frac L2\g^\m\nabla_\m\e',\quad \g^\n\g_\m\nabla_\n\e'=0.\label{eqn:cks}
\end{align}
\end{subequations}
As is well-known, the solutions of \eqref{eqn:ckv} and \eqref{eqn:cks} are just the conformal Killing vector (CKV) and conformal Killing spinor (CKS), respectively. When the metric is given as the light-cone flat metric \eqref{eqn:light-cone-metric}, the solution of the above conformal Killing conditions \eqref{eqn:conformal-killing-condition} is simply
\begin{subequations}\label{eqn:CKS-CKV-light-cone}
\begin{align}
&	\x^-=\x^-(x^-),\quad\x^{+}=\x^+(x^+),\quad\s=-\frac L2(\pa_-\x^-+\pa_+\x^+),\quad  \l_{ab}=\frac12\e_{ab}(\pa_-\x^--\pa_+\x^+),\label{eqn:CKV-light-cone}\\
& \e=\begin{pmatrix}
\tilde \e(x^-) \\ 0
\end{pmatrix},\quad \h=-L\begin{pmatrix}
0 \\ \pa_-\tilde \e(x^-)
\end{pmatrix},\quad  \e'=\begin{pmatrix}
0 \\ \tilde \e'(x^+)
\end{pmatrix},\quad \h'=-L\begin{pmatrix}
\pa_+\tilde \e'(x^+) \\ 0
\end{pmatrix}.
\end{align}
\end{subequations}
In fact, the PBH transformations with the above parameters correspond to the superconformal transformations of the dual 2D SCFT.
 
On the light-cone flat metric, Weyl invariance and super-Weyl invariance imply that
\begin{equation}
T_{-+}=T_{+-}=0,\quad S_+=S'_-=0,
\end{equation}
while the diffeomorphism and the supersymmetry Ward identity become
\begin{equation}
\pa_+ T_{--}+\pa_- T_{+-}=0,\quad \pa_+ T_{-+}+\pa_- T_{++}=0,\quad \pa_- S_+ +\pa_+ S_-=0,\quad \pa_- S'_+ +\pa_+S'_-=0,
\end{equation}
which imply that $T_{--}$ and $S_-$ are functions of $x^-$, and $T_{++}$ and $S_+$ are functions of $x^+$.

Inserting this solution into PBH transformations \eqref{eqn:PBH-stress} and \eqref{eqn:PBH-supercurrent} and using the map \eqref{eqn:bulk-boundary-constant}, we find how they transform under the asymptotic symmetries, namely
\begin{subequations}\label{eqn:conformal-trans-operators}
\begin{align}
& \d_{\x^-,\x^+} T_{--}=2\pa_-\x^-T_{--}+\x^-\pa_-T_{--}+\frac{c}{12}\pa_-^3\x^-,\label{eqn:conformal-trans-T--}\\
& \d_{\x^-,\x^+} T_{++}=2\pa_+\x^+T_{++}+\x^+\pa_+T_{++}+\frac{c}{12}\pa_+^3\x^+,\label{eqn:conformal-trans-T++}\\
& \d_{\x^-,\x^+,\tilde\e} S_-=\frac32 \pa_-\x^- S_-+\x^-\pa_- S_-+\tilde \e T_{--}+\frac c6\pa_-^2\tilde \e,\label{eqn:conformal-trans-S-}\\
& \d_{\x^-,\x^+,\tilde\e'} S'_+=\frac32 \pa_+\x^+ S'_++\x^+\pa_+ S'_+-\tilde \e' T_{++}-\frac c6\pa_+^2\tilde \e',\label{eqn:conformal-trans-S+}
\end{align}
\end{subequations}
where the first two expressions exactly render the conformal transformation of the stress-energy tensors of 2D CFT with central charge $c$. In addition, \eqref{eqn:conformal-trans-S-} and \eqref{eqn:conformal-trans-S+} imply that $S_-$ and $S_+$ are respectively primary operators of conformal dimension $(\frac32,0)$ and $(0,\frac32)$ and their supersymmetry partners are respectively $T_{--}$ and $T_{++}$, which is also in agreement with the SCFT. One can see from \eqref{eqn:conformal-trans-operators} that the modes of the stress-energy tensor and the supercurrent satisfy an $\cn=(1,1)$ super-Virasoro algebra with central charge $c$. 

An interesting remark is in order. As in the literature, we could have defined the asymptotic symmetries as the PBH transformations that do not change the metric $h_{\m\n}$ and the gravitino sources, which is weaker than \eqref{eqn:asymptotic-symmetry-condition}. However, the local Lorentz parameter $\l_{ab}$ in \eqref{eqn:CKV-light-cone} plays a crucial role in deriving \eqref{eqn:conformal-trans-S-} and \eqref{eqn:conformal-trans-S+}, which implies that it is a correct choice for the asymptotic symmetries to keep the zweibein $e^a_\m$ invariant.

The anomalous term in \eqref{eqn:conformal-trans-T--}  that is proportional to $\pa_-^3\x^-$ vanishes only when $\x^-$ is at most a quadratic polynomial in $x^-$, while the anomalous term in \eqref{eqn:conformal-trans-S-} vanishes only when $\tilde\e$ is at most a linear in $x^-$. The generators of these transformations are just $L_{0,\pm1}$ and $G_{\pm\frac12}$ that form the $\cn=(1,0)$ global superconformal algebra with no anomalous term. One can easily see that the same holds for the right-moving (or anti-holomorphic) sector.

\section{Anomaly and rigid supersymmetry}
\setcounter{equation}{0}
\label{sec:rigid-susy}

In this section, we find that no anomalous term occurs in $\cn=(1,1)$ rigid supersymmetry algebra on 2-sphere, opposite to the case of AdS$_5$/CFT$_4$ discussed in \cite{An:2017ihs,Papadimitriou:2017kzw}.

First, let us obtain the action of the rigid supersymmetry on the supercurrents. Inserting the conformal Killing spinor condition \eqref{eqn:cks} into PBH transformations \eqref{eqn:PBH-supercurrent}, we obtain the action of the rigid supersymmetry on the supercurrent, namely
\begin{subequations}\label{eqn:susy-supercurrent}
	\begin{align}
	&	[Q_\e,T_{F\m}]=-\frac14\g^\n\e(T_{\m\n}+T_{\n\m})+\frac{c}{12}\ve_{\m\n}\g^\r\nabla^\n\nabla_\r\e,\label{eqn:susy-supercurrent-L}\\
	& [Q_{\e'},T'_{F\m}]=-\frac14\g^\n\e'(T_{\m\n}+T_{\n\m})-\frac{c}{12}\ve_{\m\n}\g^\r\nabla^\n\nabla_\r\e',\label{eqn:susy-supercurrent-R}
	\end{align}
\end{subequations}
where $Q_\e\equiv \int_\cc\dd\s_\m\;\bar\e T_F^\m$ and $Q_{\e'}\equiv \int_\cc\dd\s_\m\;\bar\e' T_F'^\m$ are conserved supercharges, whose conservation follows from the CKS condition \eqref{eqn:cks} and the Ward identities \eqref{eqn:ward-supersymmetry} and \eqref{eqn:ward-supertrace}. Here $\cc$ stands for a certain Cauchy curve. 

The validity of the anomalous terms in \eqref{eqn:susy-supercurrent} can be seen as follows. Multiplying both sides of \eqref{eqn:susy-supercurrent-L} by $\g^\m$, we obtain
\begin{equation}\label{eqn:consistency-1}
\g^\m[Q_\e,T_{F\m}]=-\frac12\e (T^\m_\m+\frac{c}{12}R),
\end{equation}
where we used the relation $\Box\e=-\frac14R\e$, which follows from the CKS condition \eqref{eqn:cks}. One can readily see that \eqref{eqn:consistency-1} holds, by using super-Weyl invariance of the supercurrent operator and the conformal anomaly \eqref{eqn:anomaly-conformal}. 

Remark that \eqref{eqn:consistency-1} implies that the anomalous terms in \eqref{eqn:susy-supercurrent} cannot vanish on the non-flat background metric. It follows from \eqref{eqn:susy-supercurrent} that for any two CKSs $\e_1$ and $\e_2$, the commutator of their corresponding supercharges becomes
\begin{equation}\label{eqn:supercharge-commutator}
[Q_{\e_1},Q_{\e_2}]=-\frac14\int_\cc\dd\s_\m\;\bar\e_2\g_\n\e_1(T^{\m\n}+T^{\n\m})+\frac{c}{12}\int_\cc\dd\s_\m\;\bar\e_2\ve^{\m\n}\g^\r\nabla_\n\nabla_\r\e_1,
\end{equation}
where the c-number second term is the quantum correction due to the super-Weyl anomaly \cite{An:2017ihs,Papadimitriou:2017kzw}. A similar relation holds for the right-moving sector.

We emphasize that \eqref{eqn:supercharge-commutator}  does not imply that the rigid supersymmetry algebra on curved manifolds also gets a quantum correction due to the anomaly. In fact, the above commutator of the supercharges does not belong to the rigid supersymmetry algebra, as the vector $\bar\e_2\g_\m\e_1$ is a CKV and in general does not satisfy the Killing condition for the metric.

When the background manifold is a two-sphere, one can construct a Killing vector from the CKSs \cite{Fujii:1985bg,Lu:1998nu}. First, consider a spinor $\r\equiv \e+i\e'$ where the two spinors $\e$ and $\e'$ satisfy the relations
\begin{equation}\label{eqn:Killing-spinor-condition}
\nabla_\m\e=\frac1L\g_\m\e',\quad \nabla_\m\e'=-\frac1L\g_\m\e,
\end{equation}
which are followed by $\nabla_\m\r=-\frac iL\g_\m\r$. $\r$ is called as Killing spinor in the literature. Note that the condition \eqref{eqn:Killing-spinor-condition} can hold only when the Ricci scalar is given by $R=\frac{8}{L^2}$. Then, for any two Killing spinors $\r_1$ and $\r_2$, the vector
\begin{equation}
K^\m\equiv \bar\r_2\g^\m\r_1=\bar\e_2\g^\m\e_1-\bar\e'_2\g^\m\e'_1
\end{equation}
becomes a Killing vector.

The supercharge of $\cn=(1,1)$ rigid supersymmetry algebra on 2-sphere is just $Q_\r\equiv Q_\e+i Q_{\e'}$ associated with a Killing spinor $\r$. The commutator of two supercharges $Q_{\r_1}$ and $Q_{\r_2}$ can be inferred from \eqref{eqn:susy-supercurrent}
\begin{equation}
[Q_{\r_1},Q_{\r_2}]=-\frac14\int_\cc\dd\s_\m\;K_\n (T^{\m\n}+T^{\n\m})-\frac{c}{6L^2}\int_\cc\dd\s_\m K^\m,
\end{equation}
where the first term is a conserved charge associated with the Killing vector $K^\m$. The second term vanishes since the 1-form $K=K_\m\dd x^\m$ is Hodge dual to the exact 1-form.\footnote{This can be seen from the integrability condition of the Killing vector on two-sphere. That is, by combining $\nabla_\m K_\n=\ve_{\m\n}(\frac12\ve^{\r\s}\nabla_\r K_\s)$ and $\nabla^\m\nabla_\n K_\m=R_{\m\n}K^\m=\frac12 R K_\n$, one can obtain that $K^\m=\ve^{\m\n}R^{-1}\nabla_\n(\ve^{\r\s}\nabla_\r K_\s)$.} Explicitly, the Killing vector can be written as
\begin{equation}
	K^\m=-\frac{iL}{2}\ve^{\m\n}\nabla_\n(\bar\r_2\g_*\r_1).
\end{equation}

We thus conclude that $\cn=(1,1)$ rigid supersymmetry algebra on 2-sphere is not affected by the super-Weyl anomaly, although the transformation law of the supercurrent under the rigid supersymmetry becomes anomalous.

\section{Concluding remarks}
\label{sec:conclusions}

The main message of this note is twofold. The first is that the super-Weyl anomaly that vanishes on bosonic backgrounds is physically meaningful and in particular the super-Weyl anomaly induces a quantum correction to the transformation of the supercurrent under the rigid supersymmetry on curved manifolds. This result was claimed for the first time in \cite{An:2017ihs,Papadimitriou:2017kzw} for 4 dimensional $\cn=1$ SCFTs, and we have confirmed in this work that the same holds for 2 dimensional $\cn=(1,1)$ SCFTs.

The second is that this quantum correction induced by the super-Weyl anomaly does not affect the $\cn=(1,1)$ rigid supersymmetry algebra on two-sphere. This is opposite to the case discussed in \cite{An:2017ihs,Papadimitriou:2017kzw}, where 4 dimensional $\cn=1$ rigid supersymmetry algebra receives a quantum correction.

We emphasize that the whole procedure here can be easily extended to $\cn=(2,2)$ SCFTs. Without much effort, now one can readily see that there is no quantum correction due to an anomaly in $\cn=(2,2)$ rigid supersymmetry algebra, yet the supercurrent transforms anomalously under the rigid supersymmetry. This is important in the field theory computations (e.g. \cite{Benini2012:ui,Gerchkovitz:2014gta}) that use the supersymmetric localization technique \cite{Pestun:2007rz,Nekrasov:2002qd}. It is because the supersymmetric localization technique is based on the rigid supersymmetry algebra on curved manifolds and the assumption that there is no (rigid) supersymmetry anomaly.

\section*{Acknowledgments}

We would like to thank Gwang Il Kim for interesting discussions, as well as Un Gyong Ri for collaboration at early stage of this work. This research is supported in part by NSTC
Project No. 130-01.

\appendix

\renewcommand{\thesection}{\Alph{section}}
\renewcommand{\theequation}{\Alph{section}.\arabic{equation}}

\section*{Appendices}
\setcounter{section}{0}

\section{Conventions and ADM decomposition}
\setcounter{equation}{0}
\label{app:conventions}

We follow the conventions of \cite{Blumenhagen:2013fgp,Freedman:2012zz}. The metric signature in the bulk is $(-,+,+)$. The Greek and Latin letters refer to the coordinate and flat indices, respectively. Hatted indices ($\hat\m,\hat\n,\cdots$ and $\hat a,\hat b,\cdots$) and fields ($\hat g_{\hat\m\hat\n},\hat E^{\hat a}_{\hat\m},\cdots$) refer to the bulk indices and fields, while unhatted indices ($\m,\n,\cdots$ and $a,b,\cdots$) and fields ($g_{\m\n},E^a_\m,\cdots$) refer to the boundary indices and induced fields. Let us suppose that the bulk space is a foliation of the $r$-constant slices. We use the Majorana representation for the spinors and the $\bar\l\equiv\l^T \cc$ for the spinor $\l$ is the Majorana conjugation. Here $\cc$ is the charge conjugation matrix in 3 dimensions.

Our conventions for Gamma matrices and Levi-Civita symbols are given as follows.
\begin{align}
& \g^0=i \s_2=\begin{pmatrix} 0 & 1 \\ -1 & 0 \end{pmatrix}, \quad \g^1=\s_1=\begin{pmatrix} 0 & 1 \\ 1 & 0 \end{pmatrix},\quad  \g^2=\s_3=\begin{pmatrix}
1 & 0 \\ 0 & -1 \end{pmatrix},\\
& \e_{012}=1,\quad \e_{01}=1.
\end{align}

\subsection*{Radial ADM decomposition}

To get induced fields on radial slices, we need to decompose fields of the bulk into the radial and its normal directions, which is called the radial (ADM) decomposition. We begin with the radial-decomposition of the one-form vielbein $\hat E^{\hat a}=\hat E^{\hat a}_{\hat \m}\dd x^{\hat\m}$, and the gravitino $\j=\j_{\hat\m}\dd x^{\hat\m}$,
\bsub 
\bal 
& \hat E^{\hat a}=(Nn^{\hat a}+N^\m E^{\hat a}_\m)\dd r+E^{\hat a}_\m\dd x^\m,\\
& \j=\j_r\dd r+\j_\m\dd x^\m,
\eal
\esub 
where $N$ and $N^\m$ are called lapse and shift, respectively. The unit vector $n^{\hat a}$ is chosen to be normal to vielbeins, i.e.\begin{align} \label{eqn:unitvector-n}
n_{\hat a}E^{\hat a}_\m=0,\quad \h_{\hat a\hat b}n^{\hat a}n^{\hat b}=1,
\end{align}
such that it renders the traditional decomposition of the metric,
\be 
\dd \hat s^2=(N^2+N^\m N_\m)\dd r^2+2N_\m \dd r\dd x^\m+g_{\m\n}\dd x^\m \dd x^\n,
\ee 
where $g_{\m\n}\equiv E^\a_\m E^\b_\n\h_{\a\b}$ and $E^\a_\m$ are the induced metric and vielbein on radial slices, respectively.

The $\g$-matrices are then decomposed into
\bal
& \hat\g^r=\g^{\hat a}\hat E^r_{\hat a}=\frac 1N n_{\hat a}\g^{\hat a}=\frac 1N \g_*,\quad \g_*\equiv n_{\hat a}\g^{\hat a},\\
& \hat\g^{\m}=\g^{\hat a}\hat E^\m_{\hat a}=\g^\m-\frac{N^\m}{N}\g_*,\quad  \g^\m\equiv \g^{\hat a}E^\m_{\hat a}.
\eal
One can see that the induced $\g$-matrices satisfy the Clifford algebra on the radial slice, namely
\be 
\{\g_\m,\g_\n\}=2g_{\m\n},\quad \{\g,\g_\m\}=0,\quad \g_\m\equiv g_{\m\n}\g^\n.
\ee 
Since $\g$ commutes with all induced gamma matrices $\g^\m$, we are allowed to define the \emph{radiality} as the eigenvalue of $\gamma$ for the induced spinor fields on radial slices \cite{Martelli:2002sp,Freedman:2016yue}. The subscript $\pm$ indicates the radiality of the spinor fields, namely
\begin{equation}
	\j_\pm\equiv \frac{1\pm \g_*}{2}\j.
\end{equation}
Note that the radiality defined above coincides with the chirality in 2 dimensions.

We suppose that the flat directions $0$ and $2$ correspond to the time and radial coordinates, respectively. $n_{\hat a}$ can be arbitrary as long as it satisfies \eqref{eqn:unitvector-n}. Therefore, we choose $n_{\hat a}=(0,0,1)$ for simplicity of computations. It then follows from \eqref{eqn:unitvector-n} that $E^2_\m=0$.

$\hat\nabla$ and $\nabla$ stand for the covariant derivatives in the bulk and on the boundary respectively, while the dot symbol denotes the partial derivative with respect to the radial coordinate $r$.

Finally, we list some useful formulas. In 3 dimensions, it holds that
\begin{equation}
	\hat\g^{\hat\m\hat\n\hat\r}=-\hat\ve^{\hat\m\hat\n\hat\r},
\end{equation}
while in 2 dimensions
\begin{equation}
	\g^{\m\n}=-\ve^{\m\n}\g_*,\quad  \g_\m\g_*=\ve_{\m\n}\g^\n.
\end{equation}
Here $\ve_{\hat\m\hat\n\hat\r}$ and $\ve_{\m\n}$ stand for the totally anti-symmetric tensors in 3 dimensions and 2 dimensions, respectively.

\section{Light-cone coordinates on the boundary}
\label{app:complex-coordinate}
\setcounter{equation}{0}

In this appendix, we provide the conventions for the light-cone coordinates on the boundary. The metric is given as
\begin{equation}\label{eqn:light-cone-metric}
	\dd s^2=h_{\m\n}\dd x^\m\dd x^\n=-\dd x^+\dd x^-=-(e^0)^2+(e^1)^2,
\end{equation}
which implies that
\begin{equation}
	h_{++}=h^{++}=h_{--}=h^{--}=0,\; h_{+-}=-\frac12,\;h^{+-}=-2.
\end{equation}
Note that $x^-$ and $x^+$ are mapped to $z$ and $\bar z$ of the complex coordinates, respectively. 
The zweibein is
\begin{equation}
	e^0=\frac{1}{2}(\dd x^-+\dd x^+),\quad e^1=\frac{1}{2}(-\dd x^-+\dd x^+),
\end{equation}
or
\begin{equation}
	e^a_\m=\frac{1}{2}\begin{pmatrix}
	1 & 1 \\ -1  & 1
	\end{pmatrix}.
\end{equation}
Using this, the gamma matrices are
\begin{equation}
	\g^{+}=\begin{pmatrix}
	0 & 2 \\ 0 & 0
	\end{pmatrix},\quad \g^{-}=\begin{pmatrix}
	0 & 0\\ -2 & 0
	\end{pmatrix},
\end{equation}
and
\begin{equation}
	\ve_{-+}=e^a_- e^b_{+}\e_{ab}=\frac12.
\end{equation}

\addcontentsline{toc}{section}{References}

\nocite{*}

\bibliographystyle{jhepcap}
\bibliography{2dscft_jhep}

\end{document}